\documentclass[final,3p,times,twocolumn]{elsarticle}
\usepackage{amsmath, amssymb, amsfonts}
\allowdisplaybreaks[4]

\newcommand{\gsim}{\raisebox{-0.07cm   }{$\, \stackrel{>}{{\scriptstyle\sim}}\, $}}
\journal{Nucl. Phys. B (Proc. Suppl.)}

\begin{document}
\begin{frontmatter}
\title{{\footnotesize DESY 14--152, DO--TH 14/20, SFB/CPP--14--67 , LPN 14--110}\\
3-loop Massive $O(T_F^2)$ Contributions to the DIS Operator Matrix Element $A_{gg}$}

\author[risc]{J. Ablinger}
\ead{jablinge@risc.uni-linz.ac.at}

\author[desy]{J. Bl\"umlein}
\ead{Johannes.Bluemlein@desy.de}

\author[desy]{A. De Freitas}
\ead{abilio.de.freitas@desy.de}

\author[desy,risc]{A. Hasselhuhn}
\ead{alexander.hasselhuhn@desy.de}

\author[mainz]{A. von Manteuffel}
\ead{manteuffel@uni-mainz.de}

\author[desy,risc]{M. Round\corref{corauth1}}
\ead{mark.round@desy.de}
\cortext[corauth1]{Speaker}

\author[risc]{C. Schneider}
\ead{cschneid@risc.uni-linz.ac.at}

\address[desy]{Deutsches Elektronen--Synchrotron, DESY, Platanenallee 6, D-15738 Zeuthen, Germany}
\address[risc]{Institute for Symbolic Computation (RISC), Johannes Kepler University, Altenbergerstra{\ss}e 69, A-4040 Linz, Austria }
\address[mainz]{PRISMA Cluster of Excellence, Institute of Physics, J. Gutenberg University Mainz, D-55099 Mainz, Germany}

\begin{abstract}
Contributions to heavy flavour transition matrix elements in the variable flavour number scheme are considered at 
3-loop order.  In particular a calculation of the diagrams with two equal masses that contribute to the massive 
operator matrix element $A_{gg,Q}^{(3)}$ is performed. In the Mellin space result one finds finite nested binomial 
sums.  In $x$-space these sums correspond to iterated integrals over an alphabet containing also square-root valued 
letters.
\end{abstract}

\end{frontmatter}

\section{Introduction}

\noindent
Perturbative QCD contributions to the deep-inelastic structure functions play an important role in understanding the structure of the 
proton; to measure $\alpha_s(M_Z^2)$ \cite{Bethke:2011tr}, the parton distribution functions \cite{PDF}, and the mass of the charm quark 
$m_c$ \cite{Alekhin:2012vu} at high precision. This applies to the data analyses at HERA \cite{HERA} and future measurements at planned 
$ep$-facilities like the EIC \cite{EIC} and LHeC \cite{AbelleiraFernandez:2012cc}. At the present accuracy the QCD corrections both for 
the light and heavy flavours are requested to be known to 3-loop order. Feynman diagrams with two massive lines begin contributing to the 
structure functions at 3-loops.  Both diagrams with two equal and two unequal masses contribute. Although this work focuses only on the equal mass diagrams 
there has also been progress on the unequal mass diagrams~\cite{BW14}.  One may work in the asymptotic regime of large virtuality $Q^2 
\gg m^2$. Here $Q^2$ denotes the virtuality of the exchanged gauge boson 
and $m$ is  the heavy quark mass.  For the structure function $F_2(x,Q^2)$ it is known that the asymptotic regime holds to percent-level 
accuracy 
at NLO and for scales ${Q^2}/{m^2} \gsim 10$~\cite{Buza:1995ie}.

The heavy flavour Wilson coefficients are known to factorise into light flavour coefficients $C_{i,(2,L)}$ and process independent operator 
matrix elements (OMEs) $A_{ij}$, see~\cite{Buza:1995ie,Buza:1996wv} for the corresponding relations. The light flavour Wilson coefficients
have been calculated to 3-loop order in Ref.~\cite{Vermaseren:2005qc}.

For the equal mass diagrams that contribute to $F_2(x,Q^2)$ a set of fixed Mellin moments has been calculated~\cite{Bierenbaum:2009mv} in 
the asymptotic regime.  Meanwhile the complete set of logarithmic contributions to $F_2(x,Q^2)$ is known~\cite{Behring:2014eya}  
and for a significant list of  operator matrix elements (OMEs) and Wilson coefficients; $L_{q,2}^{(3),\rm PS}, 
L_{g,2}^{(3)}$~\cite{Behring:2014eya, Ablinger:2010ty}, 
$L_{q,2}^{(3),\rm NS}$~\cite{Ablinger:2014vwa}, $H_{q,2}^{(3),\rm PS}$~\cite{Ablinger:2014nga}, 
$A_{qg,Q}^{(3)}, A_{qq,Q}^{(3), \rm PS}$~\cite{Ablinger:2010ty}, $A_{gq,Q}^{(3)}$~\cite{Ablinger:2014lka},
$A_{qq,Q}^{(3), \rm NS}$~\cite{Ablinger:2014vwa}, and $A_{Qq}^{(3), \rm PS}$~\cite{Ablinger:2014nga} the result at general values of the 
Mellin variable $N$ has been computed in complete form.\footnote{For the notation see~\cite{Bierenbaum:2009mv}.}

In various calculations one may want to treat heavy quarks as light flavours at large enough energies. This is particularly the case 
for a series of reactions at high energy hadron colliders. The corresponding transition is described in the variable flavour number scheme
(VFNS), decoupling one heavy quark at the time~\cite{Buza:1996wv, Bierenbaum:2009mv, Bierenbaum:2009zt}\footnote{The value 
of the decoupling scale in the VFNS is process dependent and is usually {\it not} the heavy quark's
mass~\cite{Blumlein:1998sh}.}. Since the masses of the charm and bottom quarks form no strong hierarchy because ${m_c^2}/{m_b^2} \sim 
{1}/{10}$, one rather has to decouple them together. This is needed in particular for processes, where $m_c$ and $m_b$ contribute in the 
same Feynman diagram. Here a generalization of the VFNS is needed~\cite{BW14}. 

In this proceedings recent results on contributions to OMEs containing two massive fermion lines of equal
quark masses are given. In Section~\ref{secn:TF2contrib} we discuss the contributions of $O(\alpha_s^3 T_F^2 C_F)$ and $O(\alpha_s^3 T_F^2 
C_A)$ to the OME $A_{gg,Q}$ with the restriction of equal fermion masses.  Section~\ref{secn:conc} contains the conclusions. 

\section{The \boldmath{$O(\alpha_s^3 T_F^2C_{F,A})$} Contributions to $A_{gg,Q}$\label{secn:TF2contrib}}

\noindent
The contributions to the OME $A_{gg,Q}$ proportional to $\alpha_s^3 T_F^2 C_{F,A}$ were given in~\cite{Ablinger:2014uka}.  Following the 
conventions of~\cite{Bierenbaum:2009mv,Klein:2009ig} $A_{gg,Q}$ is defined as the expectation value, 
\begin{align}
A_{gg,Q} &= \langle g | O_g| g \rangle,\\
O_g &= 2i^{N-2} \textbf{S}\textrm{Sp}(F_{\mu_1, \alpha}D_{\mu_2}\cdots D_{\mu_{N-1}}F^\alpha_{\mu_N})-\textrm{trace terms},
\end{align}
with the external gluon lines on-shell. $\textbf{S}$ represents the symmetrisation with respect to the Lorentz indices and $\textrm{Sp}$ 
is the trace over colour indices.  
For further details see~\cite{Bierenbaum:2009mv,Klein:2009ig} and references therein.  For completeness the colour factors are, 
$C_F = {(N_c^2-1)}/{2N_c}, C_A=N_c$ and  $T_F={1}/{2}$ in an $SU(N_c)$ gauge theory and $N_c=3$ in QCD.  

In total 12 unique diagrams contribute to the $\alpha_s^3T_F^2C_{F,A}$ contribution of interest.  Most of the diagrams were computed 
directly. The momentum integration was performed by  introducing Feynman parameters. Then a Mellin-Barnes representation was applied to 
the 
structures that arise. These representations requested specific ways to close the contour depending on other variables involved. 
Calculating each Mellin-Barnes integral by the residue theorem led to a large number of nested sums.  These sums were handled with 
the summation technologies encoded in the package \texttt{Sigma}~\cite{SIG1,SIG2}, which uses advanced symbolic summation algorithms in the setting of difference 
fields~\cite{Karr:81,Schneider:01,Schneider:05a,Schneider:07d,
Schneider:08c,Schneider:10a,Schneider:10b,Schneider:10c, Schneider:13b}.  In addition the packages {\tt EvaluateMultiSums}, {\tt 
SumProduction}~\cite{Ablinger:2010pb, Blumlein:2012hg, Schneider:2013zna}, and  {\tt $\rho$SUM} \cite{RHOSUM} which are all 
based on {\tt Sigma}, were used. 

The remaining diagrams were solved using the integration-by-parts routines in 
{\tt Reduze2}~\cite{Studerus:2009ye, vonManteuffel:2012np}\footnote{The package {\tt Reduze2} uses {\tt Fermat} \cite{FERMAT} and {\tt
GiNac} \cite{Bauer:2000cp}.} and calculating the corresponding master integrals using differential equations and 
also applying Mellin-Barnes techniques.

Of the main results in~\cite{Ablinger:2014uka}, here the constant part of the unrenormalised OME is given.  If the unrenormalised OME corresponding to $A_{gg,Q}$ is denoted $\hat{\hat{A}}_{gg,Q}$ then one may expand in the coupling,
\begin{equation}
\hat{\hat{A}}_{gg,Q} = \frac{1}{2}[1+(-1)^N]\left\lbrace 1+ \sum_{k=1}^\infty a_s^k \hat{\hat{A}}_{gg,Q}^{(k)} \right\rbrace,
\end{equation}
for $a_s ={\alpha_s}/{4\pi}$.  The contribution at 3-loop order reads,
\begin{equation}
\hat{\hat{A}}_{gg,Q}^{(3)} = \frac{1}{\varepsilon^3} a_{gg,Q}^{(3,0)}+ \frac{1}{\varepsilon^2} a_{gg,Q}^{(3,1)}+ \frac{1}{\varepsilon}a_{gg,Q}^{(3,2)}  +   a_{gg,Q}^{(3)}.
\end{equation}
Here $\varepsilon = D - 4$ denotes the dimensional parameter.
One of our new results is the coefficient of $T_F^2$ in $a_{gg,Q}^{(3)}$.  Expressed in Mellin space it is,
{\small
\begin{align}
\label{eq:agg}
& a_{gg,Q;T_F^2}^{(3)}(N) = \nonumber\\ 
&C_F T_F^2
\Biggl\{
 \frac{16}{27} F S_1^3
+\frac{16 Q_4}{27 (N-1) N^3 (N+1)^3 (N+2)} S_1^2\nonumber\\
&+\Biggl[
-\frac{32 Q_{10}}{81 (N-1) N^4 (N+1)^4 (N+2) (2 N-3) (2 N-1)} \nonumber\\
&-\frac{16}{3} F S_2 \Biggr] S_1
-\frac{16 Q_4}{9 (N-1) N^3 (N+1)^3 (N+2)} S_2\nonumber\\
&-\frac{2 Q_{13}}{243 (N-1) N^5 (N+1)^5 (N+2) (2 N-3) (2 N-1)}\nonumber\\
&- F \left[ \frac{352}{27}   S_3- \frac{64}{3}  S_{2,1} \right]\nonumber\\
&+\Biggl[\frac{16}{3} F S_1
-\frac{8 Q_8}{9 (N-1) N^3 (N+1)^3 (N+2)}\Biggr] \zeta_2\nonumber\\
&+\frac{Q_3}{9 (N-1) N^2 (N+1)^2 (N+2)} \zeta_3\nonumber\\
&-\binom{2N}{N} 
\times \left(\sum_{i=1}^N \frac{4^i S_1(i-1)}{i^2 \binom{2i}{i}} - 7 \zeta_3\right) \nonumber\\
&\frac{16 Q_5}{3(N-1) N (N+1)^2 (N+2) (2 N-3) (2 N-1)} \frac{1}{4^N}\Biggr\}\nonumber\\
&+{C_A T_F^2} 
\Biggl\{
-\frac{4 Q_2}{135 (N-1) N^2 (N+1)^2 (N+2)} S_1^2\nonumber\\
&+\frac{16 \big(4 N^3+4 N^2-7 N+1\big)}{15 (N-1) N (N+1)} [S_{2,1} - S_3]\nonumber\\
&+\frac{Q_{12}}{3645(N-1) N^4 (N+1)^4 (N+2) (2 N-3) (2 N-1)} \nonumber\\
&-\frac{8 Q_{11}}{3645 (N-1) N^3 (N+1)^3 (N+2) (2 N-3) (2 N-1)} S_1\nonumber\\
&+\frac{4 Q_7}{135 (N-1) N^2 (N+1)^2 (N+2)} S_2\nonumber\\
&-\binom{2N}{N} \frac{1}{4^N}  \left(\sum_{i=1}^N \frac{4^i S_1(i-1)}{i^2 \binom{2i}{i}} -7 \zeta_3\right)  \nonumber\\
& \times \frac{4 Q_9}{45 (N-1) N (N+1)^2 (N+2) (2 N-3) (2 N-1)} \nonumber\\
&+\Biggl[\frac{4 Q_6}{27 (N-1) N^2 (N+1)^2 (N+2)}-\frac{560}{27} S_1\Biggr] \zeta_2\nonumber\\
&+\Biggl[-\frac{7 Q_1}{270 (N-1) N (N+1) (N+2)}
-\frac{1120}{27} S_1 \Biggr] \zeta_3 \Biggr\},
\nonumber\\ 
\end{align}}
which uses,
\begin{eqnarray}
F(N) = \frac{(2+N+N^2)^2}{(N-1) N^2 (N+1)^2 (N+2)} \equiv F.
\end{eqnarray}
Secondly, the $Q_i$ denote polynomials in $N$, see~\cite{Ablinger:2014uka} for expressions.  Thirdly, the $T_F^2$ contribution is 
described using harmonic sums \cite{Blumlein:1998if, Vermaseren:1998uu}  $S_{\vec{a}}\equiv S_{\vec{a}}(N)$,
\begin{align}
S_{b,\vec{a}}(N) &= \sum_{k = 1}^N \frac{({\rm sign}(b))^k}{k^{|b|}} S_{\vec{a}}(k), \\
 S_\emptyset &= 1,\quad b, a_i \in \mathbb{Z} \setminus \lbrace 0\rbrace.\nonumber
\end{align}
In addition to the harmonic sums there is an inverse binomial type-sum  in the $T_F^2$ contribution~\cite{Ablinger:2014bra}\footnote{For 
recent surveys on the mathematical structures occurring in zero- and single scale higher order calculations see \cite{MATH}.},
\begin{equation}\label{eq:binomsum}
\frac{1}{4^N}\begin{pmatrix}
2N\\N
\end{pmatrix} \sum_{i=1}^N \frac{4^i S_1(i-1)}{i^2 \binom{2i}{i}}  .
\end{equation}
Harmonic sums are related linearly to harmonic polylogarithms $H_{\vec{a}}(x)$~\cite{Remiddi:1999ew} in $x$-space by the Mellin transform.
They can be represented as the following iterated integrals,
\begin{align}
H(b,\vec{a};x) &= \int_0^x dt\frac{H(\vec{a};t)}{t-b},\\
H(\emptyset;x)&=1, \quad a,b \in \lbrace -1,0,1\rbrace .\nonumber
\end{align}
To include \eqref{eq:binomsum} as the transform of an iterated integral, one must extend to generalised multiple
 polylogarithms~\cite{Ablinger:2014bra, Ablinger:2013cf} including root-valued letters. For our purposes the 
following letter is needed,
\begin{align}
\mathsf{w}_9 &= \frac{1}{(1-t)\sqrt{t}},
\end{align}
see~\cite{Ablinger:2014bra} for further details.

If the Mellin transform is denoted by,
\begin{equation}
\mathsf{M}[f(x)](N) = \int_0^1 dx x^N f(x),
\end{equation}
then in $x$-space \eqref{eq:binomsum} can be built by applying the Mellin convolution to,
\begin{align}
\frac{1}{4^N}\begin{pmatrix}
2N\\N
\end{pmatrix} &=\frac{1}{\pi}\mathsf{M}\left[ \frac{1}{\sqrt{x(1-x)}}\right] ,\\
\sum_{i=1}^N \frac{4^i S_1(i-1)}{i^2 \binom{2i}{i}}  &= 
\mathsf{M}\left[ \left( \frac{H(\mathsf{w}_9,0;1-x}{x-1}\right)_+  \right] \nonumber\\
&+\mathsf{M}\left[ \left( \frac{H(\mathsf{w}_9,1;1-x}{x-1}\right)_+  \right] \nonumber\\
&+ 2 \log (2) \mathsf{M}\left[ \left( \frac{H(\mathsf{w}_9;1-x}{x-1}\right)_+  \right] ,
\end{align}
using the $+$-distribution.   Thus \eqref{eq:binomsum} is an iterated integral over the usual harmonic polylogarithm alphabet extended by 
square-root valued letters.  This is the first time such root valued letters have been found in 3-loop Wilson coefficients for 
deep-inelastic scattering.

Further results, including the renormalised $T_F^2$ contribution, can be found in~\cite{Ablinger:2014uka}.

\section{Conclusions\label{secn:conc}}

\noindent
The contributions proportional to $T_F^2 C_{F,A}$ for the gluonic massive operator matrix element at 3-loop order were calculated.  These are the diagrams with two massive lines of equal mass. 

To calculate the diagrams Mellin-Barnes integrals were used leading to nested sums.  These sums were expressed in terms of harmonic 
sums and their generalisations using automated computer algebra techniques.  Both in intermediatary and final results there are nested 
finite 
binomial sums, weighted by harmonic sums.  Moving to $x$-space the nested finite binomial sums become generalised multiple polylogarithms with square-root valued letters.

\section*{Acknowledgements}
\noindent
We would like to thank F.~Wi\ss{}brock for discussions.
This work was supported in part by DFG Sonderforschungsbereich Transregio 9, Computergest\"utzte Theoretische Teilchenphysik, 
the Austrian Science Fund (FWF) grants P20347-N18 and SFB F50 (F5009-N15), the European Commission through contract PITN-GA-2010-264564 ({LHCPhenoNet}) and PITN-GA-2012-316704 ({HIGGSTOOLS}), by the Research Center ``Elementary Forces and Mathematical Foundations (EMG)'' of J Gutenberg University Mainz and DFG, and by FP7 ERC Starting Grant  257638 PAGAP.

%-------------------------------------------------------------------------------------------------------
%\bibliography{bibliography}

\end{document}